\PassOptionsToPackage{dvipsnames}{xcolor}
\documentclass[10pt,sigconf,nonacm,screen]{acmart}

\usepackage[all]{nowidow}
\usepackage{xcolor}
\usepackage{subcaption}
\usepackage{hyperref}
\usepackage{acronym}
\usepackage{pdfpages}

\usepackage[capitalise,noabbrev]{cleveref}
\crefformat{section}{\S#2#1#3} 
\crefformat{subsection}{\S#2#1#3}
\crefformat{subsubsection}{\S#2#1#3}
\crefrangeformat{section}{\S#3#1#4 to~\S#5#2#6}
\crefrangeformat{subsection}{\S#3#1#4 to~\S#5#2#6}
\crefrangeformat{subsubsection}{\S#3#1#4 to~\S#5#2#6}

\begin{document}

\includepdf[pages=-, scale=1.0]{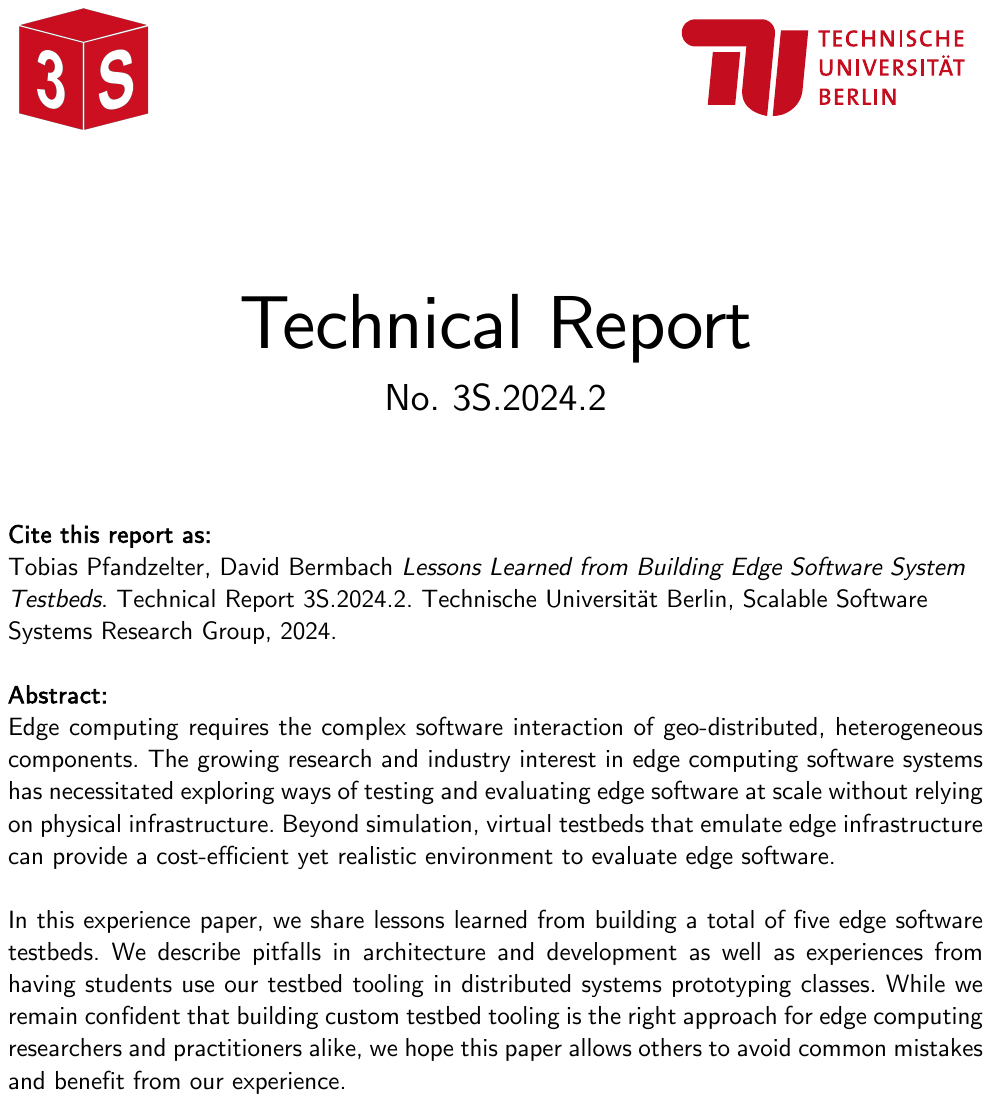}

\author{Tobias Pfandzelter}
\affiliation{%
    \institution{TU Berlin \& ECDF}
    \department{Scalable Software Systems Research Group}
    \city{Berlin}
    \country{Germany}}
\email{tp@3s.tu-berlin.de}
\orcid{0000-0002-7868-8613}

\author{David Bermbach}
\affiliation{%
    \institution{TU Berlin \& ECDF}
    \department{Scalable Software Systems Research Group}
    \city{Berlin}
    \country{Germany}}
\email{db@3s.tu-berlin.de}
\orcid{0000-0002-7524-3256}

\title[Lessons Learned from Building Edge Software System Testbeds]{Lessons Learned from Building Edge\\Software System Testbeds}

\begin{abstract}
    Edge computing requires the complex software interaction of geo-distributed, heterogeneous components.
    The growing research and industry interest in edge computing software systems has necessitated exploring ways of testing and evaluating edge software at scale without relying on physical infrastructure.
    Beyond simulation, virtual testbeds that emulate edge infrastructure can provide a cost-efficient yet realistic environment to evaluate edge software.

    In this experience paper, we share lessons learned from building a total of five edge software testbeds.
    We describe pitfalls in architecture and development as well as experiences from having students use our testbed tooling in distributed systems prototyping classes.
    While we remain confident that building custom testbed tooling is the right approach for edge computing researchers and practitioners alike, we hope this paper allows others to avoid common mistakes and benefit from our experience.
\end{abstract}

\maketitle

\section{Introduction}
\label{sec:introduction}

Edge computing extends the infinite compute and storage resources of centralized cloud data centers with on-demand computational power closer to end users and devices at the `edge' of the network~\cite{bonomi2012fog, shi2016promise,paper_bermbach2017_fog_vision}.
Researching and building edge software systems, such as IoT and eHealth applications or software abstractions and platforms, requires managing the complexity of geo-distributed, heterogeneous infrastructure connected over an unreliable network.
It is thus vital that such software systems are tested and evaluated thoroughly, in both industry and academia.

Unlike cloud services, actual edge infrastructure is, however, difficult to come by.
Instead, virtual edge system testbeds that emulate physical edge infrastructure without the need to actually deploy geo-distributed resources offer a cost-efficient alternative for software evaluation.
Researchers have proposed a number of edge testbed tools, such as \emph{EmuFog}~\cite{emufog}, \emph{Kollaps}~\cite{kollaps}, and \emph{Fogbed}~\cite{fogbed} that target different use cases with different feature sets.
We have developed five iterations of edge testbed tooling over a period of six years, namely three versions of the \emph{MockFog}~\cite{paper_hasenburg2019_mockfog,paper_hasenburg2021_mockfog2} toolkit and two versions of the \emph{Celestial}~\cite{paper_pfandzelter2022_celestial,techreport_pfandzelter2022_celestial_extended,pfandzelter2023celestialdemo} testbed for low-Earth orbit (LEO) edge computing.
We found the development of such testbeds to be a rewarding experience, as (a)~there are often no open-source testbed tools available that are both well maintained \emph{and} have the required feature set, and (b)~developing a custom testbed for edge infrastructure has improved our understanding of edge software.

Nevertheless, developing a custom edge testbed from scratch is no easy feat and there are a number of pitfalls in selecting the right tools for network emulation, deciding on virtualization and isolation techniques, or choosing the right abstractions for users.
In this paper, we thus share our experience from building five edge computing testbeds in order to give future researchers and practitioners a head start.
Specifically, we first give an overview of the history of our testbed tooling in \cref{sec:testbeds}, describing how each development cycle has influenced the next.
We then provide a selection of lessons learned in \cref{sec:lessons}, discussing pitfalls and experiences in testbed development.
Finally, we discuss avenues for future work in \cref{sec:futurework}.

\section{Background}
\label{sec:background}

We first set the context of our work by introducing the concept of edge computing (\cref{sec:background:edge}), describing the common architectural components of an edge testbed (\cref{sec:background:testbeds}), and reviewing existing testbed tools (\cref{sec:background:relwork}).

\subsection{Edge Computing}
\label{sec:background:edge}

Cloud computing promises on-demand access to infinite compute and storage resources with pay-as-you-go billing and has been broadly adopted in industry.
Centralized cloud datacenters, however, are unsuitable for applications with low latency requirements, for services that need stable, high-bandwidth connections to remote and mobile clients, or for software that stores privacy-sensitive data.
Edge computing extends the cloud by bringing resources closer to users and end devices, from servers located at radio access towers and IoT gateways to on-device computing, enabling novel application domains such as connected and automated driving, the IoT, eHealth, or VR/AR and metaverses~\cite{bonomi2012fog, shi2016promise,paper_bermbach2017_fog_vision}.
The combined edge and cloud computing paradigms are also regularly referred to as \emph{fog computing}, \emph{mist computing}, \emph{edge-to-cloud computing}, the \emph{cloud continuum}, or \emph{multi-access edge computing}.

Software running on edge computing infrastructure must manage the complex interaction of heterogeneous resources, geo-distributed clients and compute components, client mobility, and varying network performance.
This has led to a number of proposals for edge computing abstractions, such as serverless compute platforms~\cite{aslanpour2021serverless,paper_pfandzelter2020_tinyfaas,raith2023serverless} or data management middlewares~\cite{mayer2017fogstore,gupta2018fogstore,poster_hasenburg2020_towards_fbase,techreport_hasenburg2019_fbase,pfandzelter2023fred}.

Beyond cellular and IoT networks, sixth-generation (6G) mobile networks will also integrate non-terrestrial networks such as low-Earth orbit (LEO) satellite constellations~\cite{bhattacherjee2018gearing,bhattacherjee2021towards}.
Researchers have thus proposed to extend edge computing to space as well, equipping LEO network satellites with compute resources~\cite{bhattacherjee2020orbit,pfandzelter2023serverless}.
The resulting \emph{LEO edge} could provide global, low-latency access to compute resources for remote and rural applications but requires novel mechanisms to deploy and orchestrate edge software in a highly dynamic edge network~\cite{bhosale2020toward,paper_pfandzelter2021_LEO_serverless}.

\subsection{Edge Testbed Architecture}
\label{sec:background:testbeds}

\begin{figure}
    \centering
    \includegraphics[width=\linewidth]{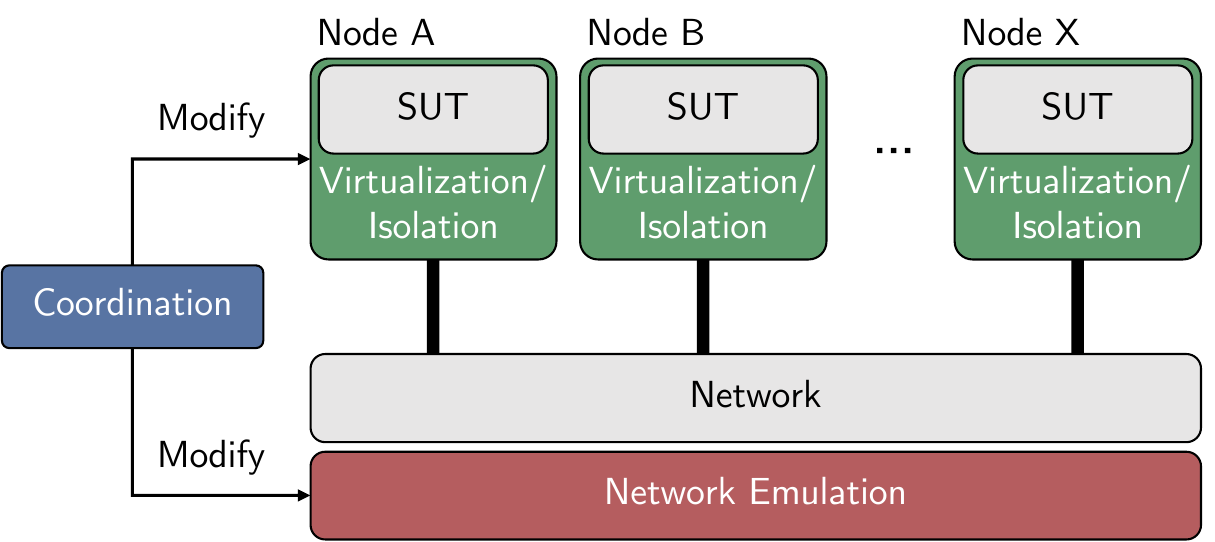}
    \caption{Edge testbeds commonly deploy each infrastructure node in its own isolation or virtualization environment and network them with emulated network characteristics. A coordinating component modifies node and network capabilities. The system-under-test (SUT), i.e., the software systems, can then run on each virtual edge node.}
    \label{fig:architecture}
\end{figure}

The salient edge infrastructure constraints are node characteristics, e.g., number of vCPUs, available storage space, or memory, and network link characteristics, e.g., link latency, bandwidth limits, or link packet loss.
As such, edge testbeds usually follow an architecture as the one shown in \cref{fig:architecture}.
Each node is emulated using a virtualization or isolation framework, such as Docker containers or cloud virtual machines, to which each node's software is deployed.
All nodes are connected using a dedicated network on which certain network characteristics are emulated using a network emulation toolkit, most frequently NetEm with Linux Traffic Control (\texttt{tc})~\cite{hemminger2005network,brown2006traffic,keller2006manual}.
Node and network characteristics are managed and modified by some coordinating entity based on a configuration file, traces, or simulation.

\subsection{Related Approaches}
\label{sec:background:relwork}

Today, there are multiple edge testbeds available beyond MockFog and Celestial, which our discussion in this paper focuses on.
In this section, we briefly discuss examples of such testbeds.

Mayer et al.~present \emph{EmuFog}~\cite{emufog}, a fog computing topology emulator based on the \emph{MaxiNet}~\cite{lantz2010network} (an extension of \emph{MiniNet}~\cite{wette2014maxinet}).
EmuFog allows deploying Docker-based applications on virtual fog nodes connected in an emulated network and is available as open-source software.
EmuFog also includes an application placement approach.

\emph{Kollaps}~\cite{kollaps} is a decentralized network emulator that scales to thousands of processes.
Kollaps can orchestrate Docker containers directly or through an orchestration tool such as Kubernetes.
\emph{Fogbed}~\cite{fogbed}, \emph{piFogBed}~\cite{xu2019pifogbed}, and \emph{Fogify}~\cite{symeonides2020fogify} are further tools for fog and edge topology emulation.

\emph{H\'ector}~\cite{hector} is a framework for the automated testing of data-intensive IoT applications that can generate custom virtual testbeds.
H\'ector is able to automatically deploy dataflow platforms such as Apache Flink or Apache Kafka and lets emulated nodes communicate using physical infrastructure.
Similarly, \emph{WoTBench}~\cite{wotbench1,wotbench2} provides a test-harness for web-of-things devices, emulating each device with a Docker container and integrating the \emph{Pumba} chaos testing tool.

\section{Testbeds}
\label{sec:testbeds}
In this section, we give an overview of all five of our edge emulation testbeds which we have developed since 2018 and specifically discuss how each of them influenced the design of the next one.
\subsection{MockFog}

The original version of MockFog\footnote{\url{https://github.com/OpenFogStack/MockFog-Meta}}~\cite{paper_hasenburg2019_mockfog} was implemented by a group of students as part of a semester project in 2018.
This version allowed users to build their desired fog infrastructure in a graphical user interface in a web browser by creating \emph{nodes} (virtual machines) and \emph{links} between them.
Users were able to specify node attributes such as images, compute resources, or available storage, as well as link characteristics such as link latency.
The MockFog \emph{node manager} would then automatically deploy a cloud virtual machine for each node to AWS EC2 or an OpenStack installation based on the infrastructure model.
A \emph{node agent} on each machine configured NetEm to emulate network links between machines.
Users could make modifications to machine and link characteristics while the emulated system was running, making, e.g., failure testing possible.

\subsection{MockFog Light}

After the release of the original version of MockFog in 2019, the requirement for a `headless' version of MockFog that could be instrumented and automated emerged while evaluating a research prototype of a pub/sub system~\cite{paper_hasenburg2020_broadcast_groups}.
We developed \emph{MockFog Light}\footnote{\url{https://github.com/OpenFogStack/MockFogLight}} by stripping a majority of user interface components from the original MockFog.
This also allowed us to drop the graph database which was used as a backend for the frontend and also for taking infrastructure snapshots to quickly undo some action.
What remained were templating files that custom MockFog programs could convert to Ansible\footnote{\url{https://www.ansible.com/}} scripts that deployed infrastructure to AWS (support for OpenStack was removed).
We also added the capability to deploy applications to our virtual nodes as Docker containers.
This modeling and deployment process allowed easy automation in order to automatically run a batch of experiments without human intervention.
MockFog Light was also used by other researchers.

\subsection{MockFog 2.0}

\begin{figure}
    \centering
    \includegraphics[width=\linewidth]{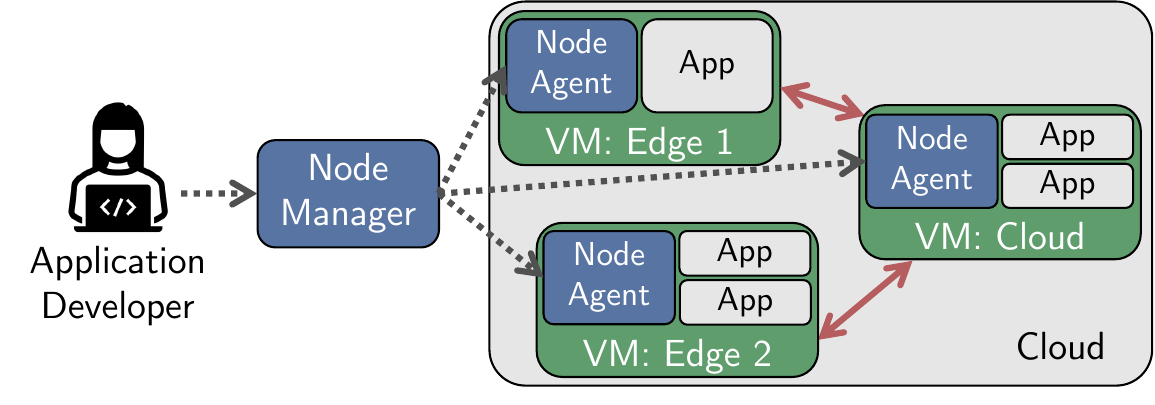}
    \caption{An example MockFog 2.0 topology where an application developer deploys a MockFog infrastructure and topology plan to the \emph{node manager}~\cite{paper_hasenburg2021_mockfog2}. Each edge and cloud node is a dedicated virtual machine running the \emph{node agent}, which translates control message from the node manager, and the application itself.}
    \label{fig:mockfog}
\end{figure}

Our goal with \emph{MockFog 2.0}\footnote{\url{https://github.com/OpenFogStack/MockFog2}}~\cite{paper_hasenburg2021_mockfog2} was to extend MockFog with support for experiment orchestration.
After two years without development on the MockFog Light research prototype, developing MockFog 2.0 in 2021 required rebuilding from the ground up in order to remove and replace outdated AWS APIs.
We show the architecture of MockFog 2.0 in \cref{fig:mockfog}.
The first new feature in MockFog 2.0 was application lifecycle management, which let users specify container images to deploy to each machine along with environment settings to automatically adapt a distributed edge application to the emulated infrastructure.
Further, the MockFog 2.0 application management module was able to retrieve experiment and application data from nodes automatically, further aiding the software evaluation process.
The second new feature was an experiment orchestrator which introduced a powerful event mechanism:
In reaction to such events, infrastructure changes could be applied, e.g., restarting a node or changing the characteristics of a network link, and application actions could be triggered, e.g., changing the configuration of the system under test or switching to a different workload.
Aside from time-triggered events, each of these changes would emit another event, thus allowing researchers to specify event chains such as \emph{``Once the preload phase has finished, start the workload of the load generator. Keep it running for five minutes, then drop network link X. Once the monitoring tool reports SLA violations, change the workload pattern of the load generator and after ten seconds double latency on all links.''}

\subsection{Celestial}

\begin{figure}
    \centering
    \includegraphics[width=\linewidth]{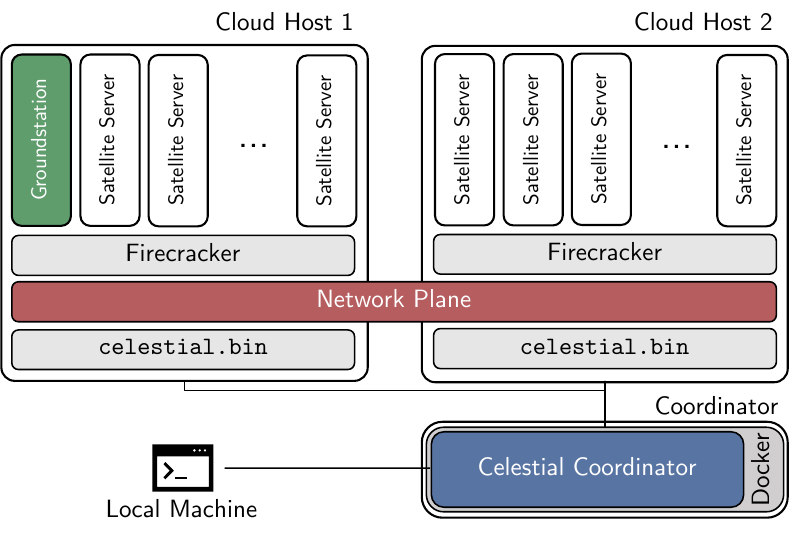}
    \caption{A Celestial emulation is based on a real-time satellite simulation at the \emph{coordinator}, which instructs \emph{hosts} on which Firecracker microVMs emulate groundstation and satellite servers. All hosts are connected with a network plane on which delays and bandwidth limits between nodes are emulated~\cite{paper_pfandzelter2022_celestial,techreport_pfandzelter2022_celestial_extended,pfandzelter2023celestialdemo}}
    \label{fig:celestial}
\end{figure}

The original version of Celestial\footnote{\url{https://github.com/OpenFogStack/celestial/tree/v0.1}}~\cite{paper_pfandzelter2022_celestial,techreport_pfandzelter2022_celestial_extended,pfandzelter2023celestialdemo} was developed shortly after MockFog 2.0 but had a different target.
Our goal was to build a highly scalable emulation platform for low-Earth orbit (LEO) edge computing, i.e., the extension of massive LEO satellite communication constellations such as SpaceX Starlink with compute resources.
LEO edge computing (a)~has more nodes than terrestrial edge computing\footnote{In practice, terrestrial edge computing will have more nodes, but research experiments can easily use less than 50 nodes to evaluate a new approach. For experimenting with LEO applications, however, researchers will usually need at least a small but full constellation of several hundred nodes.}, e.g., there are more than 4,000 satellites in the phase~\textrm{I} Starlink constellation and (b)~is more dynamic, with LEO satellites moving at speeds in excess of 27,000km/h.
Instead of using AWS EC2 cloud VMs as in MockFog, we relied on Firecracker microVMs~\cite{agache2020firecracker} for Celestial.
Firecracker microVMs, which had only recently been released at the time, are fully Linux-compatible virtual machines with a small resource footprint, allowing us to run hundreds or thousands of them on a single host.
Furthermore, it allowed us to also serve as a testbed for container orchestration or Function-as-a-Service (FaaS) platforms.
As in MockFog, we used NetEm to emulate network characteristics between VMs and targeted a deployment to cloud VMs, leveraging nested virtualization in Google Cloud.
Instead of user-defined topologies and network characteristics, Celestial relied on a satellite trajectory simulation that ran in parallel to the testbed emulation, generating ever-changing topology information in real time and continuously applying it to the testbed.
We show a complete overview of Celestial in \cref{fig:celestial}.

\subsection{Celestial v2}

Two scalability limits in Celestial prompted us to reexamine our assumptions and develop a second version of Celestial in 2023.\footnote{\url{https://github.com/OpenFogStack/celestial}}
First, we found that NetEm was unsuitable for network emulation between hundreds of machines given that it uses a tree-based filtering mechanisms that requires a linearly growing number of lookups \emph{per packet} as well as a linear increase in the latency of adding new filters.
We developed a drop-in replacement for NetEm based on eBPF that uses a hash lookup table, allowing network emulation with constant lookup times~\cite{paper_becker2022_netem}.
Second, our approach of running a network simulation concurrently with our emulated testbed was unable to scale to large networks.
Simulating thousands of satellites and calculating shortest paths between them to adjust route latency took several seconds, which was too long for a high-fidelity emulation that could capture how dynamic a satellite network is.
With Celestial v2 we decoupled a local trace generation step from the actual emulation that just replays a trace, with the added benefit of increasing repeatability of experiments.

\section{Lessons Learned}
\label{sec:lessons}

As our history of edge testbeds shows, we needed multiple years and iterations to overcome pitfalls and refine architectural decisions.
In order for other researchers, students, and developers to have a kick-start in testbed development, we share the lessons learned from building five versions of edge testbeds here.

\subsection{Merits of Building a Testbed}
\label{sec:lessons:merits}

Despite the challenges encountered along the way, the development of each testbed iteration was a rewarding experience that we would recommend to others.
Although a number of robust open-source testbed tools exist now (see also \cref{sec:background:relwork}), the relatively simple architecture of an edge testbed allows anyone to prototype a custom tool for personal use in roughly the same amount of time it takes to get comfortable with an existing one -- of course, unless advanced features such as experiment orchestration are needed.
If the goal is to evaluate an algorithm or software system on a small scale, building an automated deployment of two or three virtual machines and instrumenting NetEm on them is feasible, especially if custom deployment or instrumentation tooling is necessary.

As is common in distributed systems development, however, challenges arise once the system is scaled, as our experience with Celestial has shown.
Consequently, we here advise against building new tooling \emph{for others} to use, which provides only marginal value compared to the existing open-source testbeds.

\subsection{Specialization and Generalization}
\label{sec:lessons:specialization}

Similarly, there is a trade-off between a lean, specialized testbed that is built for a specific experiment or application and a large, general testbed tool that can emulate a variety of network topologies, application deployment strategies, or physical infrastructure backends.
The development of MockFog Light, for example, was driven by the need to run a specific experiment in an automated manner, which helped prioritize features and remove unnecessary components.
We highly recommend seeing testbeds as a means to an end rather than a standalone product that needs to cover future use-cases.

\subsection{Testbed Testing}
\label{sec:lessons:testing}

Although the development of testbeds is driven by the complexity of evaluating distributed software, testing the (distributed) testbed software itself was often not a priority for us.
The development of Celestial, for example, relied mainly on a system test that would deploy a test application on the testbed that would verify that simulated and emulated topology matched.
This was (a)~time-consuming, as each change to the code base required compiling the entire system, booting cloud virtual machines to run the testbed, and deploying the application, (b)~expensive as it required many hours of cloud instance time, and (c)~ineffective, as bugs were difficult to isolate in the system.
In hindsight, this was an obvious flaw that we addressed during the development of Celestial v2: unit and integration tests allowed us to quickly verify the functionality of smaller system components and resolve issues, which increased our development productivity.
At the same time, reducing the amount of resources to test functionality also decreased our development cost.

\subsection{Scaling Network Emulation}
\label{sec:lessons:scaling}

\begin{figure}
    \centering
    \includegraphics[width=\linewidth]{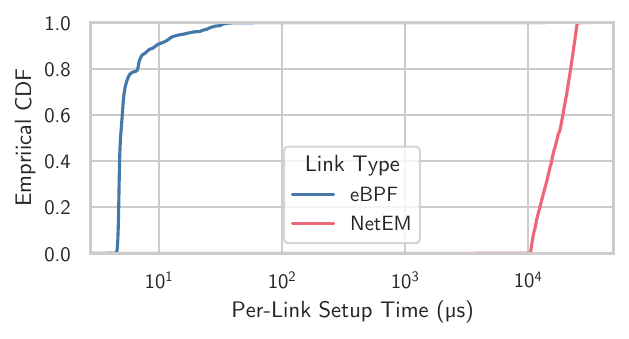}
    \caption{Setup time for $500 \times (500-1)$ links in Celestial with NetEm and our eBPF-based approach.}
    \label{fig:ebpf}
\end{figure}

One of the reasons for developing a second version of Celestial was the scalability limit of NetEm.
Although NetEm is used in all edge testbed tools we considered~\cite{herrnleben2020iot,schmidt2019constellations,becker2021edgepier,ahmed2017flexible,kollaps,xu2019pifogbed,hector,emufog,wotbench1,wotbench2,fogbed} and can be considered the state-of-the-art in Linux network emulation, it was first released in Linux kernel 2.6.7 (June 2004)~\cite{keller2006manual}, making it more than 20 years old.
It was not meant to be used at the scale we were trying to employ it at in Celestial:
For each of the thousands of microVMs on a Celestial host, we needed to specify a network filter for each other machine.
The creation of each filter, however, (a)~requires checking all existing filters for duplicates and conflicts and (b)~locks, disabling parallel creation of filters.
In experiments, we found that the configuration of links for 1,024 hosts took three hours and 24 minutes.
Note that this already used a direct connection to \texttt{netlink} instead of spawning a new \texttt{tc} process, which is also the strategy used in the Kollaps testbed.

As we were not able to find an alternative to NetEm (the newer \emph{XNetEm}~\cite{hemminger2017xnetem} does not support emulating network latency), we built our own using eBPF~\cite{paper_becker2022_netem}.
Our program injects a small eBPF filter for outgoing packets that matches target IP addresses against an eBPF map (with constant insertion and lookup times).
It would then inject a delay to the packet by modifying the packet departure timestamp, leveraging the early departure time (EDT) Linux networking model introduced in kernel 4.20~\cite{linux2018edt}.
As shown in \cref{fig:ebpf}, integrating this system in Celestial v2 reduced the mean link setup time from 17.3ms to 7\textmu{}s for a 500-satellite testbed, bringing the total setup time down from 72 minutes to just 4.3s.

\subsection{Complexity of Bandwidth}
\label{sec:lessons:bandwidth}

\begin{figure}
    \centering
    \includegraphics[width=\linewidth]{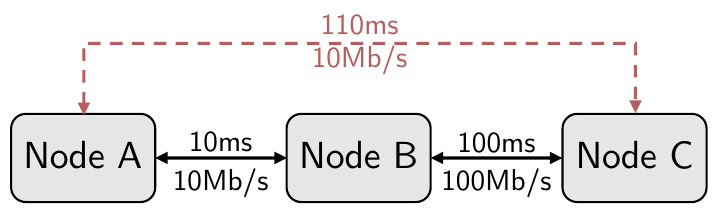}
    \caption{In MockFog and Celestial (among other testbeds), a logical topology (gray), e.g., with node A connecting to node C via node B, is reduced to a fully meshed topology. The newly introduced emulated link between A and C (shown as a dashed red line) inherits the summed latency and minimum bandwidth of the logical path between A and C.}
    \label{fig:topology}
\end{figure}

A challenge we have yet to overcome is emulating bandwidth limitations.
Both NetEm and our eBPF-based system can emulate a maximum data rate for a link between two machines.
Commonly, however, a topology as shown in \cref{fig:topology}, where machine A is connected to machine C via machine B, is reduced to a simplified topology with a full mesh between all machines (a strategy also implemented in Kollaps).
Latency for the link between A and C is simply added, while the new link's bandwidth is the minimum of the two logical links' bandwidths.
This ignores that the link is actually shared.
Kollaps implements a bandwidth monitor that adapts link rate while the testbed is running, which increases accuracy to some extent but does not solve the underlying issue.
Developing a more powerful network topology emulator is an interesting avenue for future research.

\subsection{Architecting for the Cloud}
\label{sec:lessons:cloud}

All of our testbeds were meant to be accessible to new users, especially students who, e.g., may not have access to large amounts of compute resources.
A focus of MockFog and the first version of Celestial was thus the integration of cloud infrastructure.
In MockFog, this was the direct use of AWS APIs that deployed emulated infrastructure directly to AWS EC2.
Developing MockFog 2.0 only two years after MockFog Light required updating or replacing most AWS-specific dependencies, as the cloud provider's API changed significantly in that timeframe.
Then, as AWS wound down its education program that previously allowed students to use AWS credits to experiment with cloud infrastructure, we transitioned student cloud and edge prototyping projects to Google Cloud for Education.
In 2021, a student group wanting to use MockFog 2.0 for their student project thus spent a significant amount of their project work on migrating MockFog to Google Cloud instead of working on their actual application.
While this can be educational as well, it was hardly the aim of the project and could have been avoided by providing a higher level of abstraction on top of the cloud APIs.

In Celestial, we used standard tooling that allowed deploying our system on any Linux host, whether a Google Cloud Platform, AWS, or local machine.
Nevertheless, we focused our documentation effort on Google Cloud in order to lower the barrier of entry for our students (who now had Google Cloud Education credits available) and for our own development, which was cheaper on ephemeral cloud instances compared to running it on on-site servers.
However, this raised new issues:
First, students quickly ran into limitations of Google Cloud, which limits accounts to only eight vCPUs by default, a limit that students were unable to raise.
Even with the efficiency of Celestial, this made running experiments at scale impossible.
Second, students and researchers for whom access to Google Cloud was not feasible, e.g., those that were residents in China, were without documentation on how to build and use Celestial.
While we were able to help those who asked in deploying Celestial on their infrastructure, we do not know how many were deterred from trying out Celestial.

In Celestial v2, we put an increased focus on making the documentation more accessible by removing the dependency on a cloud platform, although we still provide instructions for those looking to use Google Cloud.

\subsection{Containers and microVMs}
\label{sec:lessons:microvm}

MockFog used cloud virtual machines to isolate nodes along with Docker containers to deploy software on top of those nodes.
Virtual machines provide a high degree of isolation while Docker containers are a familiar and simple software deployment mechanism.
Most students and researchers who have used MockFog were quickly able to understand the application management process and had no issues in learning Docker or had used Docker before.

The downside of this approach was scale, as even tens of cloud virtual machines incur a significant cost.
In Celestial, moving to microVMs provided orders of magnitude of efficiency improvements by running hundreds or thousands of nodes on a single host.
A further upside was flexibility in the kind of software that could be evaluated:
While Docker containers support most software, running platform abstractions such as containers themselves was not possible.
In microVMs we can easily deploy a container management toolkit, which allows us to evaluate, e.g., the performance of container migration across edge nodes.

An unforeseen challenge in this approach was the increased barrier to usability.
Although we provide several Dockerfile-like scripts to automatically create microVM images with software dependencies, such as third-party libraries or operating system services, they lack the robustness and familiarity of the Docker tool chain.
This proved to be more difficult to use for students than anticipated.
The usability of the testbed tool is a larger factor in its design than we assumed and should be considered in any testbed design.

\subsection{Time Synchronization}
\label{sec:lessons:time}

Most cloud and web service benchmarks follow a request-response model, where clients measure the time elapsed between sending a request to the service and getting a response~\cite{book_bermbach2017_cloud_service_benchmarking,cooper2010benchmarking}.
Clients can simply measure the elapsed time using local high-resolution performance timers.
When building edge computing software to evaluate on our testbeds, however, we found that the dataflow in edge computing necessitates other time measurements.
We found that many services exhibit a source-to-sink architecture, where data generated at the edge is forwarded through the infrastructure to other edge devices or towards the cloud, often with constraints on the message latency~\cite{poster_hasenburg2018_fogexplorer,paper_hasenburg2018_supporting,paper_pfandzelter2021_zero2fog,pfandzelter2023benchmark}.
This requires a fundamentally different approach to time measurement than in the request-response model:
Each generated message or data point must be annotated with a source timestamp that can be evaluated at the sink to calculate message delivery and processing latency.
In turn, this requires accurate time synchronization between testbed machines.
In fact, time synchronization is a strength of virtualized testbeds over physical ones, as its accuracy is influenced by physical network delays.

In practice, cloud service providers offer high-fidelity time synchronization between virtual machine instances in the same availability zone.
For example, AWS EC2 integrates with the Amazon Time Sync Service that provides up to microsecond-accuracy between instances~\cite{aws2023timesync}.
Docker containers will automatically use the host clock, while in Celestial we relied on the \emph{Precision Time Protocol} (PTP) to synchronize microVM clocks with the host clock automatically~\cite{ieee1588}.

\subsection{Automation instead of GUIs}
\label{sec:lessons:automation}

\begin{figure}
    \centering
    \includegraphics[width=\linewidth]{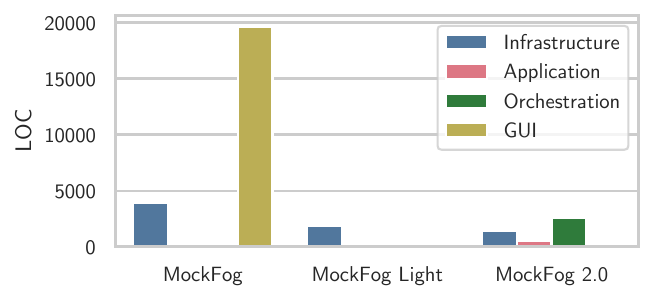}
    \caption{Counting lines of code (LOC) for each MockFog version with \texttt{scc} shows that 68\% of the original MockFog code was dedicated to the web GUI. MockFog Light reduced the codebase from 11,915 LOC to 1,953 LOC. MockFog 2.0 added new features, increasing the codebase to 4,495 LOC.}
    \label{fig:codebase}
\end{figure}

As \cref{fig:codebase} shows, the graphical user interface of the original MockFog implementation was responsible for a majority of its codebase (68\% as counted with \texttt{scc}\footnote{\url{https://github.com/boyter/scc}}).
Removing this interface in MockFog Light made MockFog development more agile, allowing us to prototype new features more quickly.
The application management and experiment control features of MockFog 2.0 allowed running an entire experiment from infrastructure configuration to data collection in a fully automated fashion, which we found to be a much more powerful interface than the original GUI.
We strongly believe that every testbed should be accessible primarily through machine-usable APIs such as a command line interface, as this enables automated software evaluation and repeatability.

Regarding the experiment orchestrator, we also realized that this hampers adoption: while it is extremely powerful in its expressiveness, it also required a steep learning curve for new users.
For many users, it was significantly easier to write a simple imperative script which made scheduled calls to the MockFog API instead of learning and then writing the descriptive experiment plan of MockFog's experiment orchestrator.

\subsection{Visualization}
\label{sec:lessons:visualization}

\begin{figure}
    \centering
    \includegraphics[width=\linewidth]{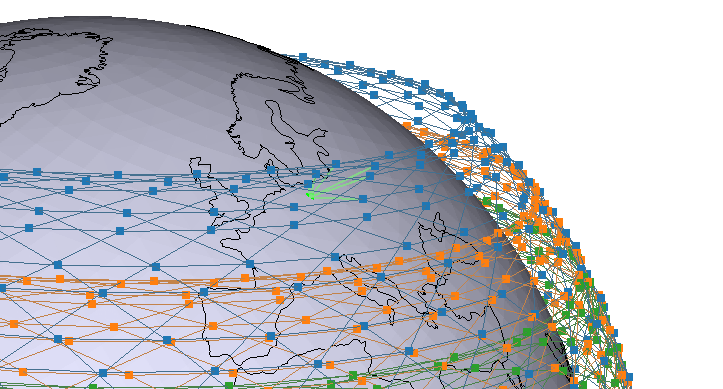}
    \caption{A screenshot of the Celestial visualization showing a ground station in Berlin connected to the proposed Amazon Kuiper satellite constellation~\cite{kassing2020exploring}.}
    \label{fig:visualization}
\end{figure}

Despite our experience with GUIs, we do recommend building a visualization interface where possible.
While this visualization should not be the primary way for a user to interact with the testbed, it should show the emulated edge infrastructure -- also for verifying correctness of configuration files.
In Celestial, we adapted the existing visualization capabilities of the \emph{SILLEO-SCNS}~\cite{kempton2021network} network simulator underlying the Celestial simulation component (shown in \cref{fig:visualization}).
This allowed new users to easily grasp the behavior of satellite infrastructure and understand how it would affect their software system.

\section{Future Work}
\label{sec:futurework}

Our experiences with edge testbed development show that there is still room for improvement in edge testbeds.
In this section, we discuss possible avenues for future work.

\subsection{Network Emulation Fabric}

In our experience, using NetEm through the \texttt{tc} command is complicated and error-prone at best, as most features of \texttt{tc} are not relevant for network emulation and any misconfiguration can lead to considerable performance degradation.
Using the \texttt{netlink} interface directly is even less user-friendly.
Beyond that, as discussed in \cref{sec:lessons:scaling}, NetEm can not be scaled to hundreds or thousands of nodes.

While our eBPF-based solution is more easily integrated in testbed tooling and provides better scalability, it still cannot solve the challenge of accurate bandwidth emulation discussed in \cref{sec:lessons:bandwidth}.
Future research work should develop a network emulation fabric that can easily be used by testbeds.
The focus here should be on scalable and accurate network emulation, including emulating bandwidth, as a building block for future testbeds rather than as its own testbed environment.

\subsection{microVM Usability}

As discussed in \cref{sec:lessons:microvm}, Docker images still provide the best programming and deployment experience for edge testbeds, as they are a familiar software packaging approach.
Even though microVMs can provide better isolation between nodes, a higher degree of infrastructure sharing, i.e., better testbed scalability, and a better feature set for edge software, we found the deployment experience of microVMs a major hurdle for users of Celestial.
We believe there is merit in further developing our idea of microVMs as testbed nodes by increasing their usability, both from a user perspective and from testbed development perspective, possibly by providing libraries for testbed developers.
For example, \emph{WeaveWorks Ignite}~\cite{ignite2023} and \emph{fly.io}~\cite{fly2024docs} already provide some level of abstraction on top of Firecracker microVMs, which could be used as guiding designs for such libraries.

\subsection{Edge Application Orchestration}

Finally, we also see the orchestration of applications at the edge as an interesting avenue for future work.
MockFog and Celestial provide some primitives for deploying applications, yet the automatic placement of services, resource allocation, request routing, and service discovery are still major challenges in running edge applications.
While these also exist in physical infrastructure and are thus not challenges specific to edge testbeds, they are a hurdle to evaluating edge software systems.
While there are existing theoretical approaches to edge orchestration~\cite{brogi2017qos,paper_pfandzelter2021_zero2fog,khare2019linearize}, few implementations exists that can be integrated easily~\cite{bartolomeo2023oakestra}.

\section{Conclusion}
\label{sec:conclusion}

In this paper, we discussed our experience from building five versions of edge computing testbeds.
The development of three versions of MockFog and two versions of Celestial has uncovered various pitfalls and interesting decisions that we believe are helpful for researchers and practitioners looking to build emulated testbeds for geo-distributed software systems.

\begin{acks}
    We thank Dr.~Jonathan Hasenburg for his insightful comments on this paper.
\end{acks}
\balance

\bibliographystyle{ACM-Reference-Format}
\bibliography{bibliography.bib}

\end{document}